\documentclass[twocolumn]{aastex62} 
\usepackage{amsmath,amstext}
\usepackage{stackengine}
 \usepackage{mathtools}
\usepackage{graphicx}
\usepackage{txfonts}

\newcommand{\noop}[1]{}

\newcommand{\ME}{M$_\Earth$\,}
\newcommand{\RE}{R$_\Earth$\,}

\newcommand{\be}{\begin{eqnarray}}
\newcommand{\ee}{\end{eqnarray}}

\newcommand{\MSun} {\mbox{$M_{\odot}$}}

\graphicspath{{./}{figures/}}

\shorttitle{Most planets may have more than 5 Myr of time to form}
\shortauthors{Pfalzner  et al.}
\begin{document}

\title{Most planets might have more than 5 Myr of time to form}

\author[0000-0002-5003-4714]{Susanne Pfalzner} 
\affiliation{J\"ulich Supercomputing Center, Forschungszentrum J\"ulich, 52428 J\"ulich, Germany}
\affiliation{Max-Planck-Institut f\"ur Radioastronomie, Auf dem H\"ugel 69, 53121 Bonn, Germany}

\author{Shahrzad Dehghani} 
\affiliation{J\"ulich Supercomputing Center, Forschungszentrum J\"ulich, 52428 J\"ulich, Germany}
\affiliation{Department of Physics, University of Cologne, Cologne, Germany}

\author[0000-0003-4099-9026]{Arnaud Michel} 
\affiliation{Department of Physics, Engineering Physics and Astronomy, Queen’s University, Kingston, ON, K7L 3N6, Canada}


\email{s.pfalzner@fz-juelich.de}


\begin{abstract}
The lifetime of protoplanetary disks is a crucial parameter for planet formation research. Observations of disk fractions in star clusters imply median disk lifetimes of 1 -- 3 Myr. This very short disk lifetime calls for planet formation to occur extremely rapidly.  
We show that young, distant clusters ($\leq$ 5 Myr, $>$ 200 pc) often dominate these types of studies. Such clusters frequently suffer from limiting magnitudes leading to an over-representation of high-mass stars. As high-mass stars disperse their disks earlier, the derived disk lifetimes apply best to high-mass stars rather than low-mass stars. Including only nearby clusters  ($<$ 200 pc) minimizes the effect of limiting magnitude. In this case, the median disk lifetime of low-mass stars is with 5 -- 10 Myr, thus much longer than often claimed. The longer timescales provide  planets ample time to form. How high-mass stars form planets so much faster than low-mass stars is the next grand challenges.
\end{abstract}

\keywords{circumstellar matter, protoplanetary disks, open clusters and associations, planet formation}

\section{Introduction}
\label{sec:intro}
The disks surrounding young stars provide the building material for planets. While terrestrial mantle rocks show that the Earth took tens of Myr to form \citep{Halliday:2006}, such direct formation dating is impossible for gas giants and exoplanets. Therefore, the frequency of disks around stars of different ages is used as a method to obtain information about the time available for planet formation \citep[][]{Haisch:2001}. The derived median disk lifetime is a "make-or-break" test for planet formation theories. As stellar ages of individual stars are intrinsically highly imprecise \citep[e.g.,][]{Bell:2013,Richert:2018}, disk fractions, $f_d$,  in young star clusters are used instead. As clusters consist of fairly coeval stars, their ages can be determined with higher accuracy than for individual stars.

The decline of the disk fraction with cluster age, $t$, has been shown for many different disk indicators, such as infrared excess or accretion signatures \citep{Haisch:2001,Hernandez:2007,Fedele:2010,Ribas:2014,Richert:2018}.
Exponential fits of the form $f_d(t) = exp(-t/\tau)$ provide an inconsistent picture with median disc lifetimes ranging from  \mbox{1 -- 3.5 Myr} to  \mbox{5 -- 10 Myr}. 
As at least half the stars in the field seem to harbour planets \citep[e.g., ][]{Winn:2015}, the shorter disk lifetimes would imply extremely rapid planet growth.

\begin{table*}[t]
\caption{Examples of the fraction of old and nearby clusters in disk lifetimes studies}
\label{tab:old_cluster_number}
\centering
      \begin{tabular}{lccccllllll}
  \hline \hline
        \multicolumn{1}{l}{Reference}    & \multicolumn{1}{l}{Number of clusters} & \multicolumn{2}{l}{ Fraction of clusters with}& \multicolumn{1}{l}{disc lifetime} \\
        
        \multicolumn{1}{l}{}    & \multicolumn{1}{l}{} & \multicolumn{1}{l}{$t_c\geq$}5 Myr & \multicolumn{1}{c}{$d\leq$200 pc} & \\
     \hline
     \citet[][]{Richert:2018}   &69  & 0.00 & 0.00 & 1.3 -- 3.5 \\
     \citet[][]{Haisch:2001}    & 7  & 0.15 & 0.28 & 3.0  \\
     \citet[][]{Hernandez:2008} & 18 & 0.17 & 0.22 & $<$ 5   \\
     \citet[][]{Fedele:2010}    & 10 & 0.20 & 0.33 & 2.9 \\
     \citet[][]{Briceno:2019}   & 9  & 0.55 & 0.00 & 2  \\
     \citet[][]{Ribas:2014}     & 22 & 0.15 & 0.54 & 2  \\    
     \citet[][]{Ribas:2015}     & 11 & 0.5  & 0.73 & 5.2 \\
     \citet[][]{Michel:2021}    &11 & 0.36 & 0.91 & 8.0 \\
     this work, Fig. 1 middle   &14 & 0.50 & 1.0 & 7 -- 8 \\
\hline
           \hline    
      \end{tabular}
 
\end{table*}

%

The uncertainties in cluster age determination \citep[][]{Bell:2013} is a known problem in deducing disk lifetimes from cluster disk fractions. Besides,  environmental effects can lead to lower disc fractions in the dense cluster centres \citep[][]{Guarcello:2007,Pfalzner:2014}. Also, these effects play a part, here we show that the main reason for the discrepancy in the derived disk lifetimes is its sensitivity to cluster selection regarding age and distance. We suggest that the stellar mass dependence might skew the results towards short disk lifetimes. We find that restricting the sample to nearby ($<$ 200 pc) clusters with an adequate balance between young and old clusters leads to much longer median disk lifetimes of 5--10 Myr. The disk fractions of the individual clusters in this sample are from two works \citep[][]{Michel:2021,Luhman:2021}. 
Within these two sources, the same method was used to determine the disk fractions.

\section{Correlation between cluster sample and disk lifetime}

The work by \citet[][]{Haisch:2001} held the promise that additional data would lead to a well-defined decline of the disk fraction with cluster age. However, the more data became available, the wider became the spread. This spread is often interpreted as being caused by external disk dispersal mechanisms. While this explains the lower disc fraction in dense clusters (typical $n \approx$ 10$^4$ pc$^{-3}$ \citep[][]{Gutermuth:2005}) ( see clusters indicated in green in Fig. 1, top left \citep[][]{Stolte:2015,Vincke:2018,Concha:2021}, it fails to account for the higher disc fractions in sparse clusters (typical $n <$ 0.5 pc$^{-3}$)). The black curve in Fig. 1 is based on clusters with $n <$ 600  pc$^{-3}$, where the effect of close stellar flybys on the disc fractions is restricted to small areas close to the cluster centre. Many of them have no O stars or much few than Upper Sco or UCL/LCC, excluding external photo-evaporation as the main cause for the difference in disk fraction. In total the environmental effect on the disc fraction should be small ($\ll$5\%) \citep[][]{Vincke:2018,Concha:2021}. Furthermore, the disc fractions of older sparse associations and co-moving groups are up to 25\% higher (Fig. 1, top right, red data points) higher than for the black curve. Thus  density effects cannot be dominantly responsible for this large difference. 

\begin{figure*}[t]
\includegraphics[width=0.49\textwidth]{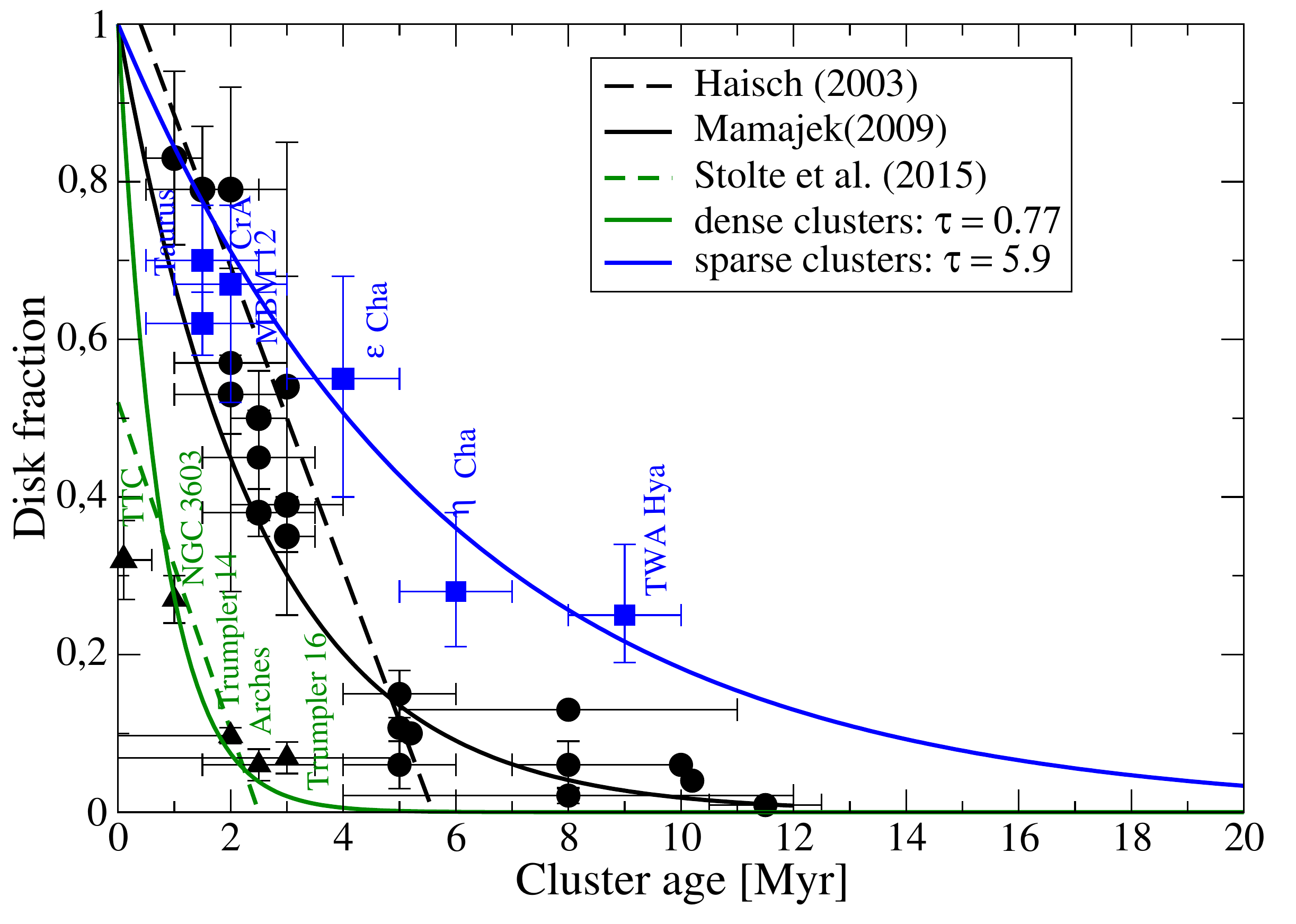}
\includegraphics[width=0.49\textwidth]{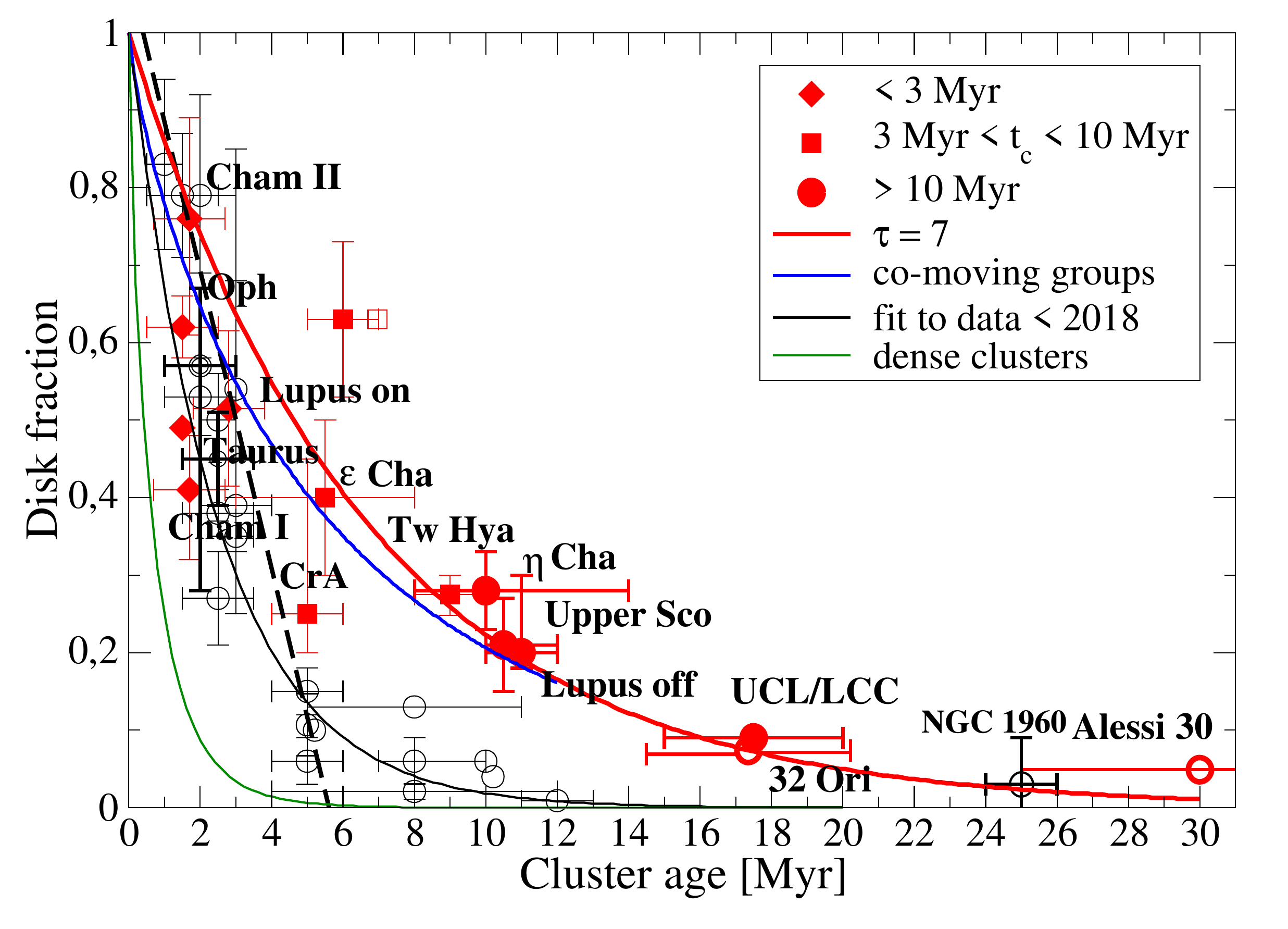}
\includegraphics[width=0.49\textwidth]{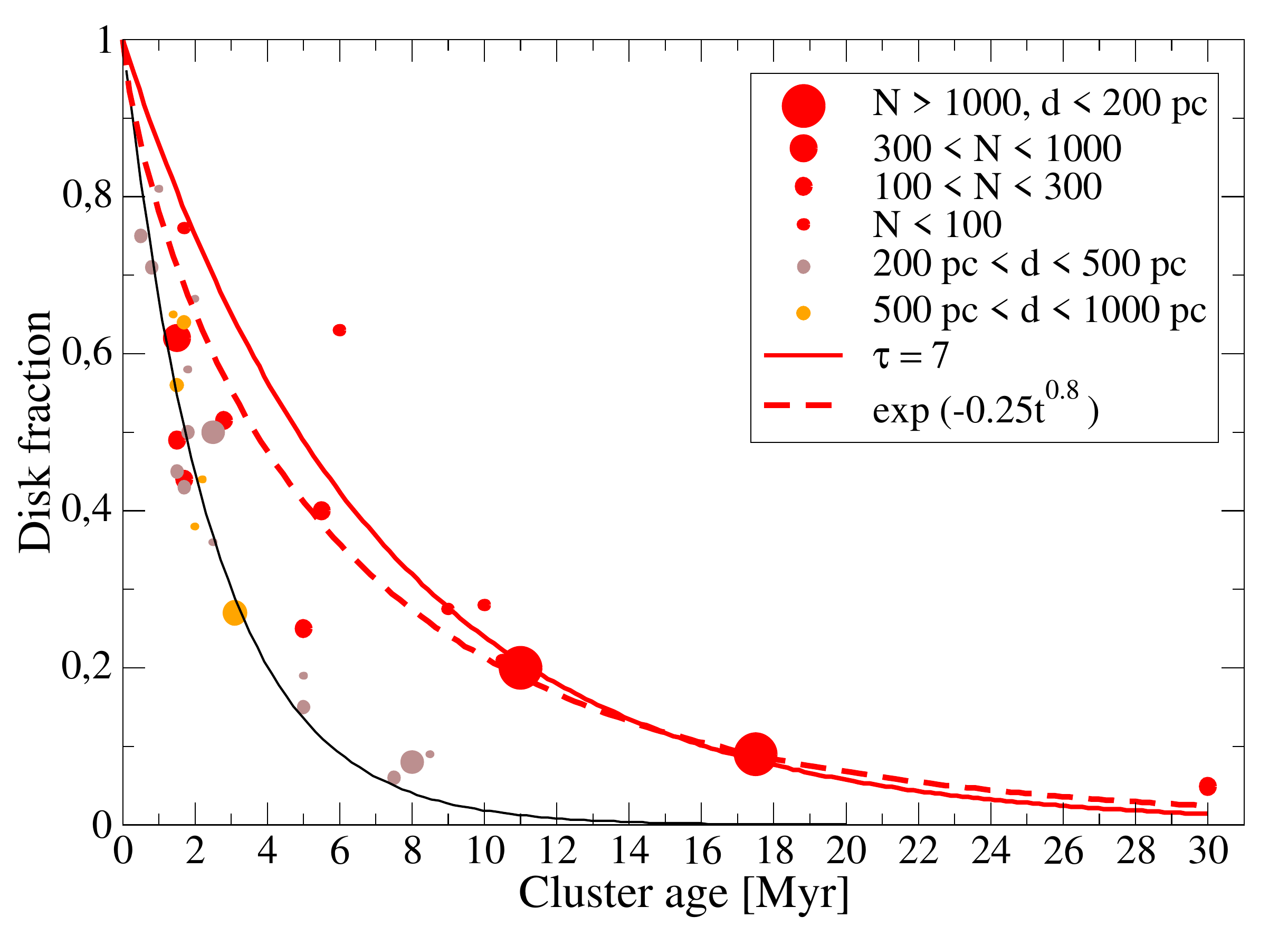}
\includegraphics[width=0.49\textwidth]{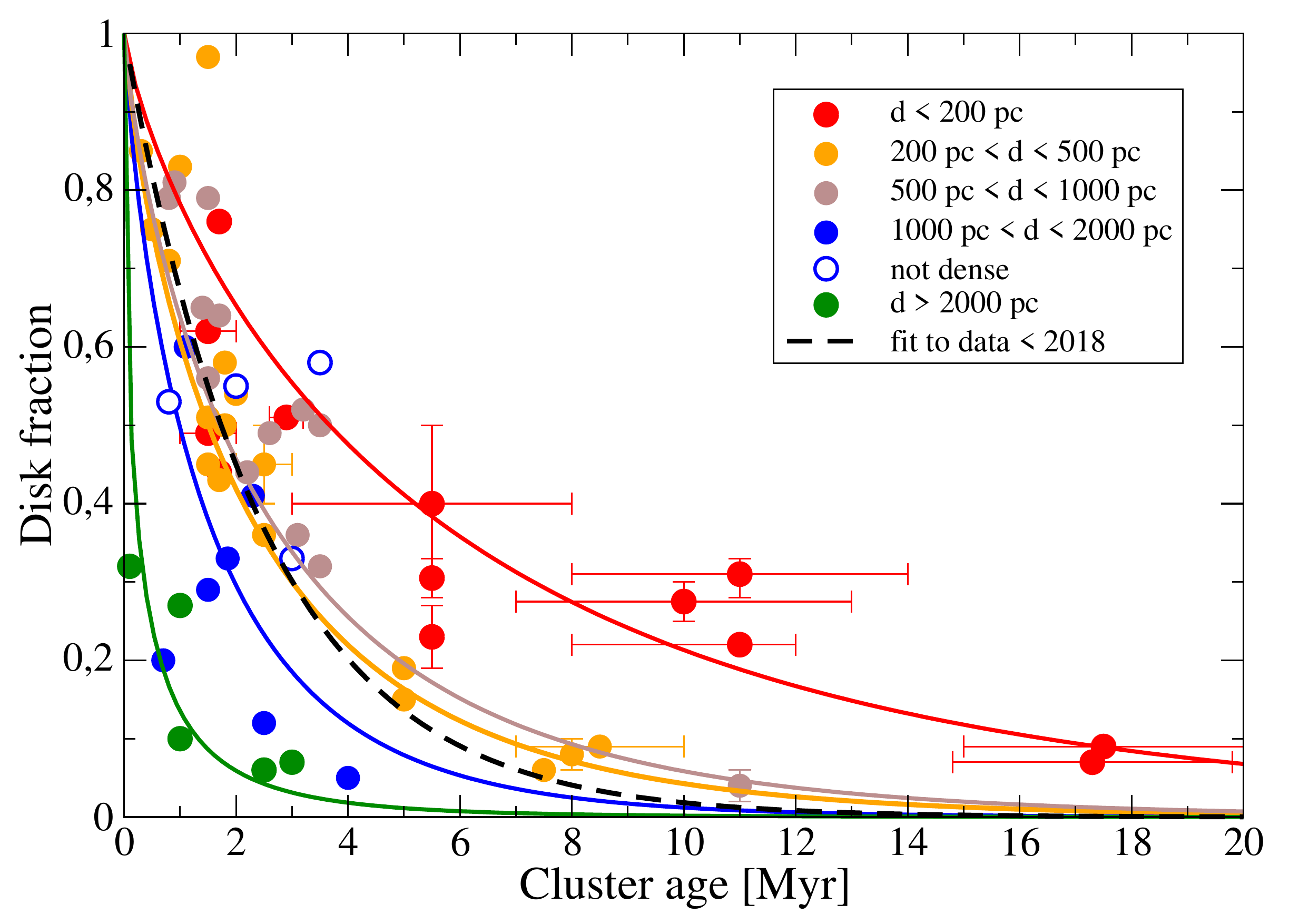}
\caption{Disk fraction vs cluster age. Top: Distinction between areas of different density. Clusters generally used are shown in black, compact, dense clusters in green, and sparse clusters in blue. The original fit by Haisch (dashed black line). Middle: Fit using only clusters within 200 pc with the age range 1 -- 20 Myr equitably covered (red line and symbols). 
Bottom: Clusters within 1000 pc. Symbols area proportional to the number of stars considered and the colour representing the cluster distance.
}
\label{fig:disc_fraction_age_old}
\end{figure*}

Most clusters are detected as over-density relative to the background and elevated infrared excesses. Young clusters are identifiable even if they contain only a few hundreds or even tens of stars. When young clusters lose most of their gas content at the end of the star formation phase \citep[for example, ][]{Lada:2003,Kuhn:2019}, their size increases to 5 -- 10 times its initial value at ages \mbox{1 -- 5 Myr}. Thus, older clusters must contain $N >$ 1000 stars to be detectable. 
This is why low-$N$ clusters older than 5 Myr are missing from plots like Fig. \ref{fig:disc_fraction_age_old} top right. Exceptions are the co-moving groups, as they are discovered as moving in the same phase space independent of their surface density.

Due to their initial compactness, clusters are identifiable at much larger distances at young ages than later. Table \ref{tab:old_cluster_number} shows that studies including a significant fraction of distant clusters arrive at shorter median disc lifetimes. Even studies that include many old clusters, but all at large distances, show this trend. 
Our central hypothesis is that limiting magnitude at large distances introduces a bias towards brighter high-mass stars. As high-mass stars tend to lose their discs earlier \citep[see, for example,][]{Carpenter:2006,Ribas:2015}, this skews the resulting disk lifetimes towards shorter values.

Most young clusters contain 10 -- 1000 times fewer stars than the older clusters \citep[][]{Pfalzner:2014}.   This makes the individual disc fraction of young clusters less statistically significant in deriving the disk lifetime than these of older clusters. In many studies, young clusters are over-represented, while older clusters are under-represented. For example,  in \citet[][]{Haisch:2001} about 85\% of clusters and all 69 clusters in \citet[][]{Richert:2018} were $\leq$ 5 Myr. Table \ref{tab:old_cluster_number} shows examples of the fraction of clusters \mbox{$>$ 5 Myr} in various studies. Studies with large fractions of young clusters tend to derive short median disc lifetimes ($<$ 5 Myr) than those with more long-lived clusters ($>$ 5 Myr).


\section{Median disk lifetimes}
\label{sec:disk_lifetime}

To avoid these biases, we include nearby clusters \mbox{($<$ 200 pc)} with an equitable weight over the entire age range  \mbox{1 -- 20 Myr.}  Our sample contains four clusters aged \mbox{1 - 3 Myr,} five aged \mbox{3 -- 8 Myr} and four aged \mbox{8 -- 20 Myr.} 
\mbox{Fig. \ref{fig:disc_fraction_age_old}} top, right shows that the resulting decline of disk fraction with cluster age (red solid line) is shallower than in studies mainly based on young distant clusters (black line)(for values  see Table \ref{tab:disc_fractions_old_new}).  We fitted different types of curves to the data. We find that the dependence of the disk fraction on cluster age can be approximated by a one-parameter exponential function of the form  ${f_D = \exp (-t/7})$ with a median disc lifetime of 5.0 Myr or two-parameter exponential function of the form $f_D = \exp (-0.25 t^{0.8})$. These fits exclude the data of the even older clusters $\sim$ 25 Myr old NGC 1960 and the 25 -- 35 Myr old Alessi 30 \citep{Galli:2021_1} as it is unclear whether their disks are protoplanetary or debris in nature. As an alternative, we also performed a Gaussian fit of the corresponding disc life distribution (see accompanying research note). In this case, we obtain a median disc lifetime of 6.5$\pm$1.5 Myr. Independent of the applied method, the derived median disk lifetimes exceed considerably the usually quoted \mbox{1 -- 3 Myr}. We emphasize that 25\% of disks exist beyond 10 Myr and $\approx$ 10\% beyond 15 to 20 Myr.

The question is how statistically significant is this result. Everything else being equal,  the disk fractions of high-$N$ clusters are statistically more significant. Fig. \ref{fig:disc_fraction_age_old} bottom left shows the disc fraction as a function of cluster age for the clusters within 1 kpc (see Table \ref{tab:disc_fractions_old_new}). However, here the symbol area is approximately proportional to the number of stars, $N$, considered in determining the disc fraction. It becomes apparent that the disc fractions of Upper Sco and UCL/LCC are considerably more statistically significant than those of the younger clusters. In total, there are 448 stars surrounded by a protoplanetary disc in Upper Sco and UCL/LLC in \citep[][]{Luhman:2021}. Therefore it is improbable that some extreme outlier stars dominate by their exceptionally long disc lifetimes. Thus the result of a median disc lifetime exceeding 5 Myr seems to be highly statistically significant.

Besides these selection effects, some older studies assumed Upper Sco was younger and had a lower disc fraction. Age estimates for Upper Sco range from 5 Myr to 12 Myr \citep[ for example, ][]{Preibisch:2002,Pecault:2016}, however, with an increasing consensus towards an age of 10 -- 12 Myr \citep[][]{Feiden:2016,Ascensio:2019}.  Before GAIA, significant uncertainties in membership existed, especially in the cluster outskirts. Disk fractions in the outer areas could be up to 3 times lower due to false positives at that time \citet{Rizzuto:2012}. One strategy to avoid the problem of false positives was to consider only the central cluster areas. However, at least for high-mass clusters older than 3 -- 5 Myr, this introduces a bias toward lower disk fractions caused by external disk destruction \citep[][]{Pfalzner:2014}. Nowadays, the false positive rate in the outskirts of clusters is much lower. Recent disk fractions of Upper Sco are nearly twice as high as the 11 \% used in several older studies. Similarly, for the 15 -- 20 Myr old UCL/LCC region, disk fraction values increased from 1\% -- 3\% to 9 \% nowadays. 

Why does restricting to nearby clusters with an even spread in cluster ages lead to a longer disc lifetime? \citet[][]{Michel:2021} found the disk fraction of Upper Sco to be very similar to that of co-moving groups of similar age. They argue that the low-UV radiation in both samples is the reason for the similarity in disk fractions. In the following, we reason that distance rather than similar radiation levels might lead to disc fractions being similar for these otherwise quite different environments.

\section{Median disk lifetime: a question of stellar mass}
\label{sec:stellar_mass}

Many studies found that disk fractions to be lower for high- than low-mass stars \citep[e.g.,][]{Carpenter:2006,Roccatagliata:2011,Yasui:2014,Ribas:2015,Richert:2018}. 
Here we hypothesise that this stellar mass dependence of disc fractions is partly responsible for the wide spread in disc fractions at any given age. Most stars are of low mass (M- and K-type); however,  observations of more distant clusters or at low sensitivity suffer from limiting magnitude. Thus mean stellar mass is higher as low-mass stars are under-represented in these samples. This bias towards higher stellar mass affects the derived disc lifetimes directly. As high-mass stars lose their disks faster\citep[for example, ][]{Ribas:2015}, distant cluster disk fractions are systematically lower. Consequently, the derived disc lifetimes are shorter.

We test this hypothesis by plotting the disk fractions as a function of stellar age colour-coded by cluster distance (Fig. \ref{fig:disc_fraction_age_old} bottom left). A strong correlation between the slope of the decline in disk fraction with cluster distance becomes apparent. Obviously, current median disk lifetimes suffer from a strong selection effect connected to cluster distance. Here cluster distance is only a crude proxy for missing out on low-mass stars in the more distant samples but is sufficient to demonstrate the problem.

The mean density of distant clusters like Arches or Trumpler 14 clusters is about a thousand times higher than in clusters like the ONC. These dense clusters are located at distances $>$ 2000 pc. Therefore they are strongly affected by limiting magnitude; The derived very low disc fractions usually apply only to stars with masses $>$ 1 \MSun. However, environmental effects should also apply to these clusters \citep[][]{Mann:2014,Vincke:2016,Ansdell:2017}. However, determining the relative importance of mass dependence vs environment remains a challenging task.

Upper Sco has a disk fraction of 5\%$^{+4\%}_{-3\%}$ for B7--K5.5-type stars \citep[][]{Luhman:2021} compared to 22\%$\pm$0.02 for low-mass stars (M3.7-M6). Similar, for the UCL/LLC, the disk fraction is 0.7\%$^{+0.06}_{-0.04}$ for higher mass stars compared to 9\%$\pm$1\% for low-mass stars. The statistical significance of these data is quite high, even when separated into mass bins. For M3.75-M6 stars (22\%) it is based on 633 objects, for K6-M3.5 stars (18\%) on 311 objects, and for B7--K5.5-type stars (5\%$^{+4\%}_{-3\%}$) on 76 objects in \citep[][]{Luhman:2021}. So even for the high-mass star, the statistical sample is as large as many young low-mass clusters' total population. The statistical significance is even higher for the UCL/LLC data, with sample sizes of 2488, 725 and 452 in the equivalent stellar mass bins.

\begin{figure}[t]
\includegraphics[width=0.50\textwidth]{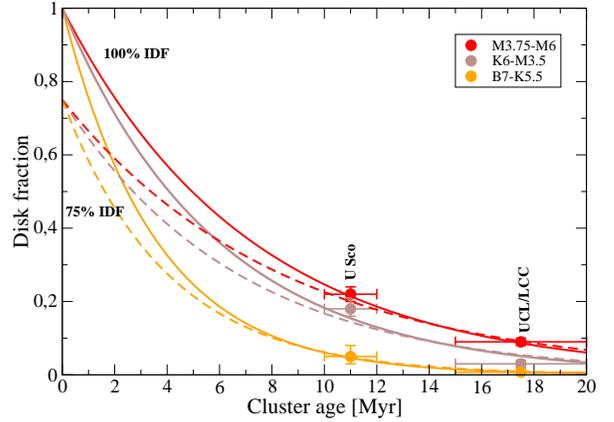}
\caption{Disk fraction vs. cluster age.  Top: colors indicating different cluster distances. Middle: colors indicating different limiting magnitudes. Bottom: Fits for mass-dependent disc lifetimes based on the high-significance values of Upper Sco and UCL/LCC assuming an initial disc fraction of 100\% (solid line) and 75\%  dashed line}.
\label{fig:disc_fractions_mass}
\end{figure}

This dependence of the disk fractions on stellar mass translates into shorter disc lifetimes for high- than low-mass stars. Figure \ref{fig:disc_fractions_mass} shows a simple extrapolation based on the  Upper Sco and UCL/LLC data, assuming an initial disk fraction of 100\%. Basing the curves on just three points is not ideal (the third one being the assumed primordial disc fraction, but as shown in Fig. \ref{fig:disc_fraction_age_old}, bottom, the fit for the total disk fraction holds for additional clusters. Nevertheless, it illustrates the critical point of the mass-dependence of the decline in disc fraction. Using Fig. \ref{fig:disc_fractions_mass} as a first indication, we expect that 50\% of the low-mass stars still have a disk at $\approx$ 5 Myr. By contrast, only $<$ 20\% of the high-mass stars retain their disks at that age. Half of the high-mass stars have already lost their disks at $\approx$ 2  Myr.   Thus the median disk lifetime of high- and low-mass stars seems to differ at least a factor of two.  

There is considerable uncertainty of the zero-age disk frequency \citep[][]{Michel:2021}. If all stars are initially surrounded by disks, the median disk frequency for the low-mass stars would be 5 -- 7 Myr, whereas for initial disk frequencies of 80\%, the median disk lifetime increases to \mbox{6 -- 10 Myr.} (see Fig. \ref{fig:disc_fractions_mass} bottom). Here again the higher values are obtained when performing a Gaussian fit the corresponding disc life distribution.

\begin{figure}[h]
\includegraphics[width=0.47\textwidth]{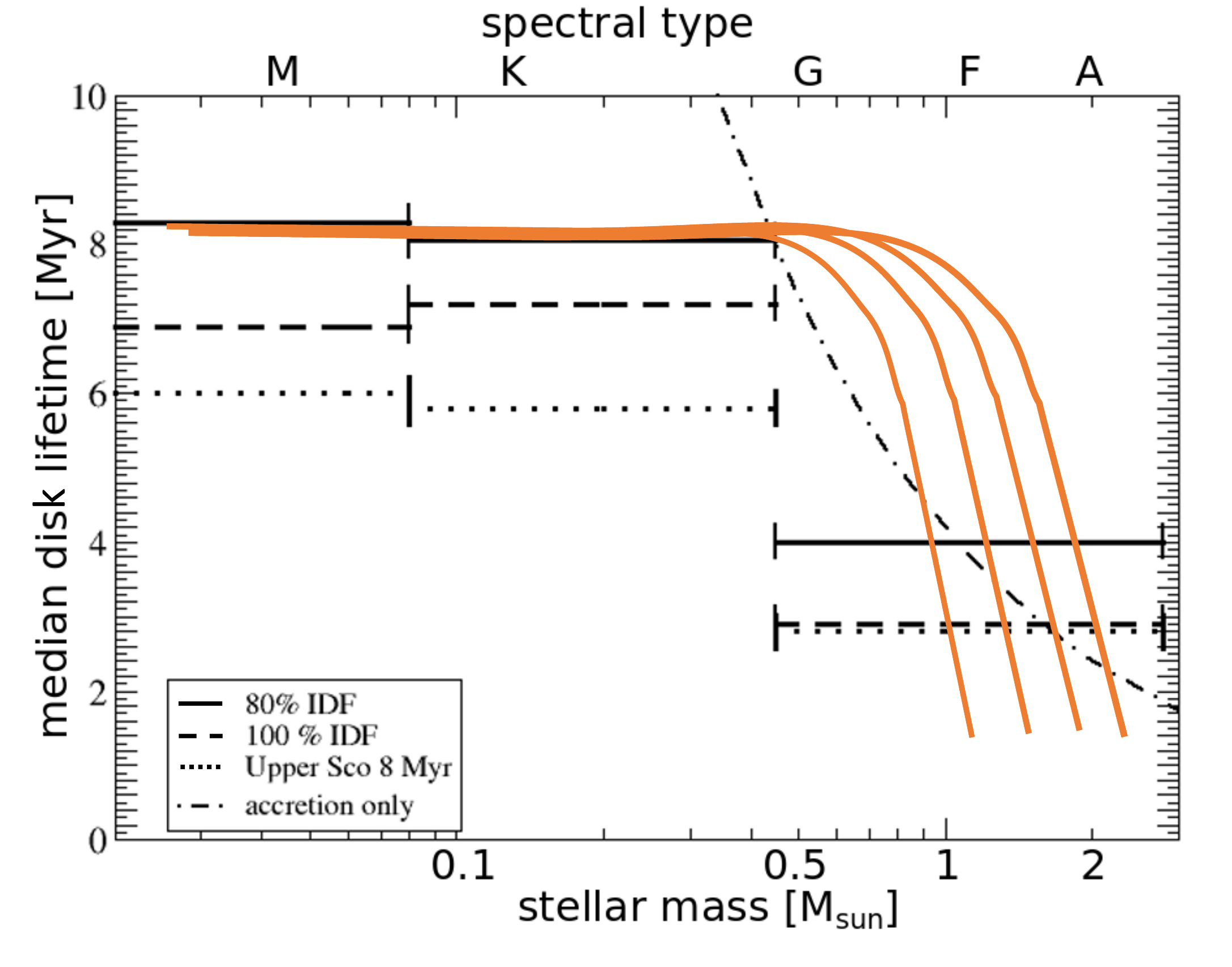}
\includegraphics[width=0.5\textwidth]{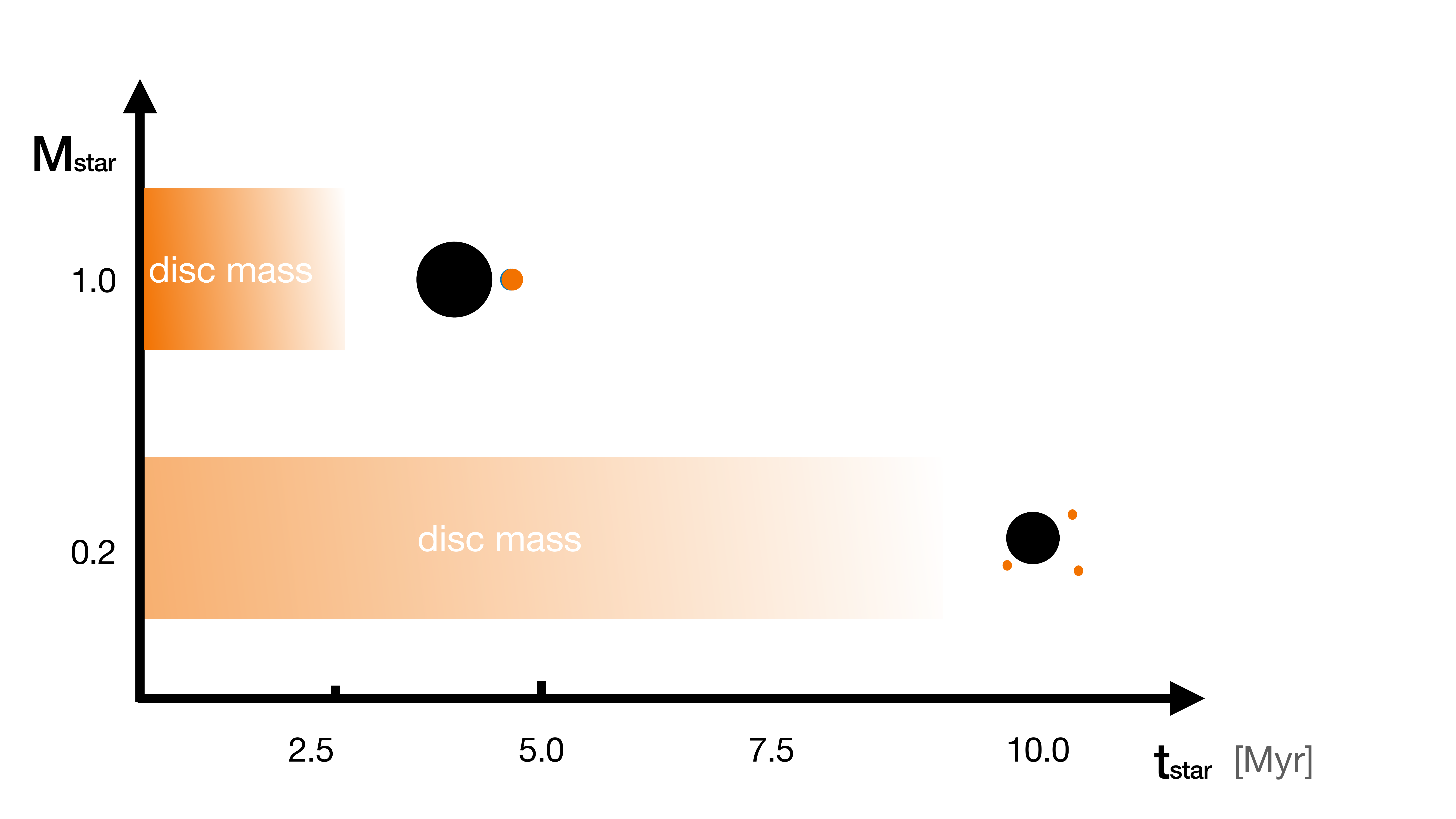}
\caption{Top: Median disk lifetime as function of stellar mass assuming 100 \% initial disc fraction (dashed lines,) 80\% initially disc fraction (solid lines) and assuming a lower Upper Sco age of 8 Myr instead of 11 Myr (dotted lines).The orange lines show the general trend including the uncertainty for higher mass stars. Bottom: Correlation between disk lifetime and properties of planetary system.}
\label{fig:mass}
\end{figure}

The mass dependence disk lifetime could be determined from the disk fractions for the different stellar masses. Unfortunately, there is a lack of statistically meaningful data for high stellar masses. Fig. \ref{fig:mass} illustrates the general trend based on the Upper Sco and UCL/LCC data. Both scenarios (initial disk frequency of 100\% and 80\%) show the same overall trend: For M -- K stars, the median disk lifetime is basically the same, but for higher mass stars, it is considerably lower. However, where precisely the decline in disk lifetimes happens remains uncertain. This relation requires urgently further observational investigation.


Why do high-mass stars lose their disks so much earlier than low-mass stars?
One answer might lie in their higher efficiency in accreting material and photo-evaporation. \citet{Martijn:2021} find that stars with masses exceeding  \mbox{0.8 \MSun\ } have shorter lifetimes due to these two effects; nevertheless, lifetimes up to 15 Myr are still possible for all host star masses up to 2 \MSun. The critical role of accretion is supported by observations finding that the lowest M stars still that retain a disk at ages $\approx$ 8 -- 10 Myr also show moderate accretion levels \citet[][]{Venuti:2019}.


\section{From the diversity of disk life times to that of planets}
\label{sec:planets}

The biggest surprise in exoplanet observations is the immense diversity in planets and planetary systems \citep[][]{Howard:2013,Gaudi:2021}. Suggested causes 
are, among others, differences in the disk mass and disk mass profile \citep[][]{Kokubo:2002}, the metallicity of the stars \citep{Petigura:2018}, the location of rings in discs \citep[][]{Marel:2021}, and the type of cluster environment \citep[][]{Bate:2018,Vincke:2018,Winter:2020}. 

The variation in individual disk lifetime has also been suggested to influence the type of planet that forms and the architecture of the planetary systems \citep[][]{Carpenter:2005,Luhman:2012, Ribas:2015}. If that holds, the stellar mass dependence of the disc lifetime should result in differences in the planets as a function of stellar mass.

Disk masses scale quasi linearly with stellar mass \citep[e.g., ][]{Andrews:2020}. Thus high-mass stars have a much larger gas and dust reservoir for planet formation. 
Therefore it is not surprising that main-sequence FGK stars host more larger planets than low-mass stars \citep[][]{Howard:2012, Sabotta:2021}. However, the situation is different for smaller planets. M dwarfs host about a factor of three more small planets (1.0--2.8 \RE) than main-sequence FGK stars \citep[for example, ][]{Mulders:2021}. However, the mass is not simply redistributed into more, smaller planets. Surprisingly, the average heavy-element mass decreases with increasing stellar mass from 7 \ME for M stars to 5 \ME for G and K stars and 4 \ME for F stars. Thus despite M star disks containing ten times less mass, they are nearly 20 times as efficient than F stars in converting the disk's heavy-element content into planetary material. The higher heavy-element content also corresponds to a higher fraction of stars with planetary systems for low-mass stars \citep[][]{Yang:2020, He:2021}. 

The strong dependence of the disk lifetime on stellar mass may explain the high planet-formation efficiency in low-mass stars. High-mass stars seem to produce their high-mass gas giants on timescales shorter than 3 Myr while failing to form additional low-mass planets. By contrast, low-mass stars form large planets to a lower degree; however, their long disk lifetimes allow for the formation of many small planets. In a way, slow but steady beats fast and short. 

Forming low-mass planets on time scales $>$ 5 Myr lowers the hurdles for the standard accretion model. However, explaining why high-mass stars form more massive planets on considerably shorter time scales remains a significant challenge. A possible explanation is that their higher disk masses make their disks more prone to gravitational instabilities. However, as $M_d / M_s \sim$ const independent of stellar mass, the parameter $Q$ that describes the stability of disks should be the same for high and low mass stars. Thus unstable disks are at least not a straightforward explanation. Thus planet formation around high-mass stars remains an open question.  
%
%


\section{Summary and Conclusion}

We investigated the role of cluster sample selection on derived median disc lifetimes. We find that samples with a large fraction of distant, young clusters \mbox{($>$ 200 pc;} $<$ 5 Myr) tend to derive short disc lifetimes (1 -- 3 Myr). Samples including higher fractions of nearby, older clusters arrive at higher disc lifetimes ($>$ 5 Myr). Restricting to clusters closer than 200 pc aged between 1 and 20 Myr, we obtain a median disk lifetime of 5 to 10 Myr. 

One main reason for this discrepancy is that distant clusters are affected by limiting magnitude. Therefore, the disc fraction of, on average, higher mass stars is determined in distant clusters. As the disk lifetime of high-mass stars is shorter than for low-mass stars, the disc fractions of more distant clusters seem lower due to this selection effect. We conclude that disc lifetimes derived from samples including many distant clusters ($>$ 200 pc) represent mostly high-mass stars. Indeed, if we only consider the high-mass stars in the sample limited to distances $<$ 200 pc, we recover a disk lifetime of only 2 -- 4 Myr.  

However, these disc lifetimes are not representative of most of the stars. Low-mass stars have a median disc lifetime of 5 -- 10 Myr. The actual value depends mainly on the assumed initial disc fraction. If all stars are surrounded by disks at cluster ages $t_c$ = 0 Myr, the median disc lifetime is  5-- 6 Myr. However, some stars seem to be born diskless or lose their disk extremely rapidly to planet formation. An initial disk fraction of \mbox{70\%-- 80 \%} would increase the disk dissipation times for low-mass stars to 8 -- 10 Myr and that of high-mass stars to 4 -- 5 Myr. 

For low-mass stars, the median disk lifetime of \mbox{5 -- 10 Myr} significantly relaxes the temporal constraints on planet formation. These long disk lifetimes allow for sufficient time for planets to form via accretion. 
The diversity of disk lifetimes might influence the structure of the emerging planetary system. It could be responsible for low-mass stars having considerably higher efficiency in using the heavy-element content in their disk for planet formation. Despite having considerably lower disk masses to start with, these low-mass stars produce a larger number of lower-mass planets. The real challenge remains to explain how high-mass stars can form planets on such a short timescale.

Currently, the effect of the environment on the disk lifetime is still not quantifiable. Only disk fractions for high-mass star clusters have been determined for dense clusters due to their general large distances ($>$ 2000 pc). A comparison is difficult even for those high-mass stars as the initial disc fraction in these dense clusters is unknown. Its uncertainty is very high, as none of the measured disc fractions exceed 32\%.

Generally, determining the initial disc fraction is the next step to determine disc lifetimes accurately. This includes potential dependencies on stellar mass, cluster density and binarity.  \\

\bigskip
We thank the referee for the constructive report and useful suggestions that have significantly improved our manuscript. We would like to thank K. Luhman for helpful advice on interpreting his results on the disk fractions in Upper Sco and UCL/LCC.

\begin{table*}
        \caption{Disk fractions}
        \centering
        \begin{longtable}{lrclllcclccccc}
        \hline
        Identification & $d$ & Age & $N_{stars}$  & $f_d$ &  Limit  & Median mass &  log($\rho_c)$ &Source  \\
                       & pc  & Myr &              &           &    & [\MSun]& [\MSun/pc$^3$] \\
                     \hline
      $ d \ll$ 200 pc\\
       \hline
       \hline%
       Alessi 30         &  108 & 30       & 162  & 0.049\footnote{possibly debris disk fraction} &     0.04 \MSun & & & a)  \\
       UCL/LLC           &   150 & 15--20  & 3665 & 0.09            &   & 0.15                               &  -0.85 -(-1.05) & b) \\    
       32 Ori       &   95  & 15--20 &  160 & 0.07 &   & 0.15  & & f1 \\
       Upp Sco           &  145 & 10--12  & 1774 & 0.22/0.20       & 0.01 \MSun  & 0.15                                &-0.59 &   b, c) \\
       Lupus -off cloud  & 160  & 10--12  &  60    & 0.21 $\pm$ 0.06 &  0.05 \MSun        &                                  & &   l)\\
       $\eta$ Cha        &   94 &  8--14  &  40    & 0.28/33         &          &                                  & &   d)\\
       \hline
       TW Hya            &   56 & 7--13   &  56   & 0.25/0.30       &          &                                   &  & d)\\
       Lupus-on cloud    &  160 & 6       &  30  & 0.63 $\pm$ 0.04 &   0.05 \MSun       &                                    & & l) \\
       CrA               &  152 & 5       & 146  & 0.23 $\pm$ 0.4 & 0.04 \MSun &                                   & & d)\\  
       $\epsilon$ Cha    &  101 & 5 (3--8)& 90& 0.5/0.3\footnote{The disk fraction is much higher in the center than the outskirts ($>$10 pc) of $\eta$ Cha}&      &                                                                                      & & d, e)\\
       \hline
       Lupus             &  158 & 2.6--3.1 & 158  & 0.50/0.53  & 0.03 \MSun &                                      & & a, d) \\
       Cham I            &  188 & 1.7      & 183  & 0.44  & 6 $<$G $<$20    &                                      & & f) \\
       Cham II           &  197 & 1.7      & 41   & 0.76  & G12-G18         &                                      & & f)  \\
       Taurus            &  128-196 & 1--2 &  137    & 0.49  &       0.05 \MSun          &                                   & & i)    \\
       Ophiuchus         &  139 & 1--2     &  420    & 0.62  &                 &                                   & & i)    \\     
        \hline  
        \hline
        200 pc $< d <$ 500 pc\\
       \hline
       \hline
        25 Orionis        &  330   &    8.5 $\pm 1.5$ & 26   & 0.09      &     &   &   & g)\\
        Ori 1a            &  355   &    8             & 811  & 0.08      &  0.1 \MSun   &   &   -0.54   & a1) \\
        $\gamma$ Vel      &  345   &    7.5           & 125  & 0.06      &  0.2 \MSun   &   &      & i) \\
        $\lambda$ Ori     &  414   &    5             & 43   & 0.19      &     &   &      & i) \\
        OriOB1b           &  414   &    5             & 278  & 0.15      &  0.04 \MSun   &   &      0.55 & i) \\
        IC 348            &  310   &   2-3            & 310  & 0.50/0.40 &   &   &     2.2 & d) \\
        $\sigma$ Ori      &  414   &   2.5            & 71   & 0.36      &     & 0.3 \MSun  &      & i) \\
        NGC2068/NGC2071   &  400   &   1-3            & 67    & 0.54     &     &   &      & j) \\
        Berkley59         &  400   &   1.8            & 201  & 0.50      &  0.1 \MSun   & 0.78  &  & k) \\
        Serpens South     &  415   &   1.8            & 26 & 0.58       &  0.1 \MSun   &  0.3 & & k) \\
        ONC Flank         &  414   &   1.7            & 236 & 0.43       &  0.13 \MSun  &  0.52 & 1.3 & k) \\
        OMC               &  414   &   1.5            & 181 & 0.45       &  0.09 \MSun  &  0.32 &  & k) \\
        L1630N            &  400   &   1.5            &     & 0.97       &     & &      & l) \\
        Lynds1641         &  400   &   1.5            & & 0.51       &     &    &  & m) \\
        NGC1333           & 320    &   1.0            & 73 & 0.81       &     &  &    & j) \\
        Flame/NGC2023     &  414   &   0.8            & 142 & 0.71       &  0.1 &  0.38 \MSun  &  & k) \\
        Serpens           &  425   &   0.5            & 137   & 0.75       &     &  &    & j) \\
        NGC2024           & 415    &  0.3             & & 0.85       &     &     & 2.424& a1)\\
        \hline   
        \hline
        500 pc $< d <$ 1000 pc \\
       \hline
       \hline
        NGC 7160        & 900      & 11 $\pm$ 1      & --    & 0.04 $\pm$ 0.03  & & &  &b1) \\
        CepOB3b-East    & 700      & 3.5             &       & 0.32             & & &  &x) \\
        CepOB3b-West    & 700      & 3.5             &       & 0.50             &  & & &y) \\
        NGC2264         & 760      & 3.2             & 324      & 0.52             & & & &c1)\\
                        & 751      & 3.1             &       & 0.38             & & & &j) \\
        Trumpler37      & 900      & 2.6             &       & 0.49             & & & &j) \\
        CepC            & 700      & 2.2             &  59   & 0.44             & 0.1 \MSun  & 0.47 && k) \\
        CepA/A          & 700      & 2.0             &  77   & 0.38             & 0.1 \MSun  & 0.43 && k) \\
        MonR2           & 830      & 1.7             &  208  & 0.64             & 0.09 \MSun & 0.47 & 1.72  &k) \\
        LkH$\alpha$101  & 510      & 1.5             &  140  & 0.56             & 0.1 \MSun  & 0.56 && k) \\
        L988e           & 700      & 1.5             &       & 0.79             &  & & & d1) \\
       \multicolumn{3}{r}{Continued on next page}\\%

        \end{longtable}
  \end{table*}

   \begin{table*}
     \renewcommand\thetable{2}
        \caption{continued}
        \centering
        \begin{longtable}{lrclllcclccccc}
        \hline
        Identification & $d$ & Age & $N_{stars}$  & $f_d$ & Limit  & Median mass& log($\rho_c)$ &Source  \\
                       & pc  & Myr &              &           &    & [\MSun]& [\MSun/pc$^3$] \\
\hline%
\hline%
        CepA/C          & 700      & 1.4             &  86   & 0.65             & 0.1 \MSun & 0.38 && k) \\
        RCW36           & 700      & 0.9             &       & 0.81             & 0.1 \MSun & 0.35 & &k) \\  
        W40             & 500      & 0.8             &       & 0.79             & 0.1 \MSun & 0.53 & &k \\
       \hline
       \hline
                             1000 pc $< d <$ 2000 pc \\
       \hline
       \hline
       NGC6231         & 1585     & 4    & & 0.05               & & & &n)  \\
       NGC2282         & 1650     & 3.5  & & 0.58               & & & &o)  \\
       NGC7129         & 1260     & 3    & & 0.33 $\pm$ 0.22    & & & &q)  \\
       W3Main          & 1950     & 3    & & 0.07               & & & &r)  \\
       NGC2244         & 1880     & 3    & 570 & 0.445              & & & 1.03 & e1) \\
       NGC2362         & 1480     & 2.5  & & 0.12               & & & &k) \\
       M8              & 1300     & 2.3  & & 0.41               & & & &k) \\
       AFGL333         & 2000     & 2.0  & & 0.55 $\pm$ 0.5     & & & &s)  \\
       NGC 6611        & 1995     & 2.0  & & 0.59                      \\
                       & 1750     & 1.2  & & 0.34               & & & 1.45 &u)\\
       Pismis 24        & 1700     & 1.85 & & 0.33               & & & &t) \\
       Cyngus OB2       & 1450     & 1.5  & & 0.29               &  &  &1.61  &u)  \\
       M 17             & 2000     & 1.1  & 35 & 0.6                & 0.12 \MSun & 3.68 & & k)  \\
       Sh2-106         & 1400     & 0.8  & 92 & 0.53 $\pm$ 0.1     & 0.13 \MSun & 0.6  & & k)  \\
       NGC 6530         & 1300     & 0.7  & & 0.20               & & &  &v)  \\
       \hline
       \hline
       $d >$ 2000 pc \\
       \hline
       \hline
       Trumpler 15      & 2360    & 8.0 & & 0.021    & 1 \MSun   & & & w)  \\
       Bochum 1         & 2800    & 5.0 & & 0.086    & A -B      & & & w) \\
       Quintuplett      & 8000    &  4.0   & 766 &  0.04    & A - B    & &3.7 & z)  \\
       Trumpler 16      & 2700    & 3.0 & & 0.069    & 1 \MSun   & & &w)  \\
       Trumpler 14      & 2800    & 2.0 & & 0.097    & 1 \MSun   & &4.3 &w)  \\
       Arches           & 8000    & 2.5 & & 0.092    & A - B     & & 5.6  &z) \\
       NGC 3603         & 3600    & 1.0 & & 0.27     &           & & 5.0  &z) \\
       TTC              & 2700    & 0.1 & & 0.32     & 1 \MSun   & & &w)\\
       \hline
       \hline
              \end{longtable}
              \tablerefs{a) \citet[][]{Galli:2021_1}, b) \citet[][]{Luhman:2021}, c) \citet{Luhman:2020}, d) \citet[][]{Michel:2021}, e) \citet[][]{Dickson:2021}, f) \citet[][]{Galli:2021_2}, g) \citet[][]{Ribas:2014}, i) \citet[][]{Manzo:2020}, j) \citet[][]{Sung:2009}  k) \citet{Richert:2018},l) \citet{Spezzi:2015}, m)\citet{Fang:2013}, n)\citet{Damiani:2016}, o)\citet[][]{Dutta:2015}, q) \citet[][]{Stelzer:2009}, r) \citet[][]{Bik:2014}, s)\citet[][]{Jose:2016}, t) \citet[][]{Fang:2012}, u) \citet[][]{Guarcello:2016}, v) \citet[][]{Damiani:2006}, w) \citet[][]{Preibisch:2011}, x) \cite{Allen:2008}, y) \citet[][]{Allen:2012}, z) \citet[][]{Stolte:2015}, a1) \citet{Briceno:2019}, b1) \citet{Hernandez:2008}, c1) \citet[][]{Sousa:2019}, d1) \citet[][]{Allen:2008}, e1) \citet[][]{Balog:2007}
              }
     \label{tab:disc_fractions_old_new}
\end{table*}
%


\newpage
\bibliographystyle{aa} 
\bibliography{references} 

%
%


\end{document}